\def\year{2019}\relax
\documentclass[letterpaper]{article} 
\usepackage{aaai19}  
\usepackage{times}  
\usepackage{helvet}  
\usepackage{courier}  
\usepackage{url}  
\usepackage{graphicx}  
\usepackage{multirow}
\usepackage{subfig}
\usepackage{amsmath,amssymb}
\usepackage{algorithm}
\usepackage{algorithmic}
\newcommand{\tabincell}[2]{\begin{tabular}{@{}#1@{}}#2\end{tabular}}
\begin{document}

\title{Coupled CycleGAN: Unsupervised Hashing Network for Cross-Modal Retrieval}
\author{Chao~Li,\textsuperscript{\rm 1} 
		Cheng~Deng,\textsuperscript{\rm 1}\thanks{Corresponding author} 
		Lei~Wang,\textsuperscript{\rm 1}
		De~Xie,\textsuperscript{\rm 1}
		Xianglong~Liu\textsuperscript{\rm 2}\thanks{Corresponding author}\\
	\textsuperscript{\rm 1}School of Electronic Engineering, Xidian University, Xi'an 710071, China\\
	\textsuperscript{\rm 2}State Key Lab of Software Development Environment, Beihang University, Beijing 100191, China\\
	li\_chao@stu.xidian.edu.cn, \{chdeng.xd, lwang.xidian, xiede.xd\}@gmail.com, xlliu@nlsde.buaa.edu.cn\\
}
\maketitle

\begin{abstract}
In recent years, hashing has attracted more and more attention owing to its superior capacity of low storage cost and high query efficiency in large-scale cross-modal retrieval. Benefiting from deep leaning, continuously compelling results in cross-modal retrieval community have been achieved. However, existing deep cross-modal hashing methods either rely on amounts of labeled information or have no ability to learn an accuracy correlation between different modalities. In this paper, we proposed \underline{U}nsupervised coupled \underline{C}ycle generative adversarial \underline{H}ashing networks~(UCH), for cross-modal retrieval, where outer-cycle network is used to learn powerful common representation, and inner-cycle network is explained to generate reliable hash codes. Specifically, our proposed UCH seamlessly couples these two networks with generative adversarial mechanism, which can be optimized simultaneously to learn representation and hash codes. Extensive experiments on three popular benchmark datasets show that the proposed UCH outperforms the state-of-the-art unsupervised cross-modal hashing methods. 
\end{abstract}
\begin{figure*}[!t]
	\begin{center}
		\includegraphics[width=0.95\textwidth]{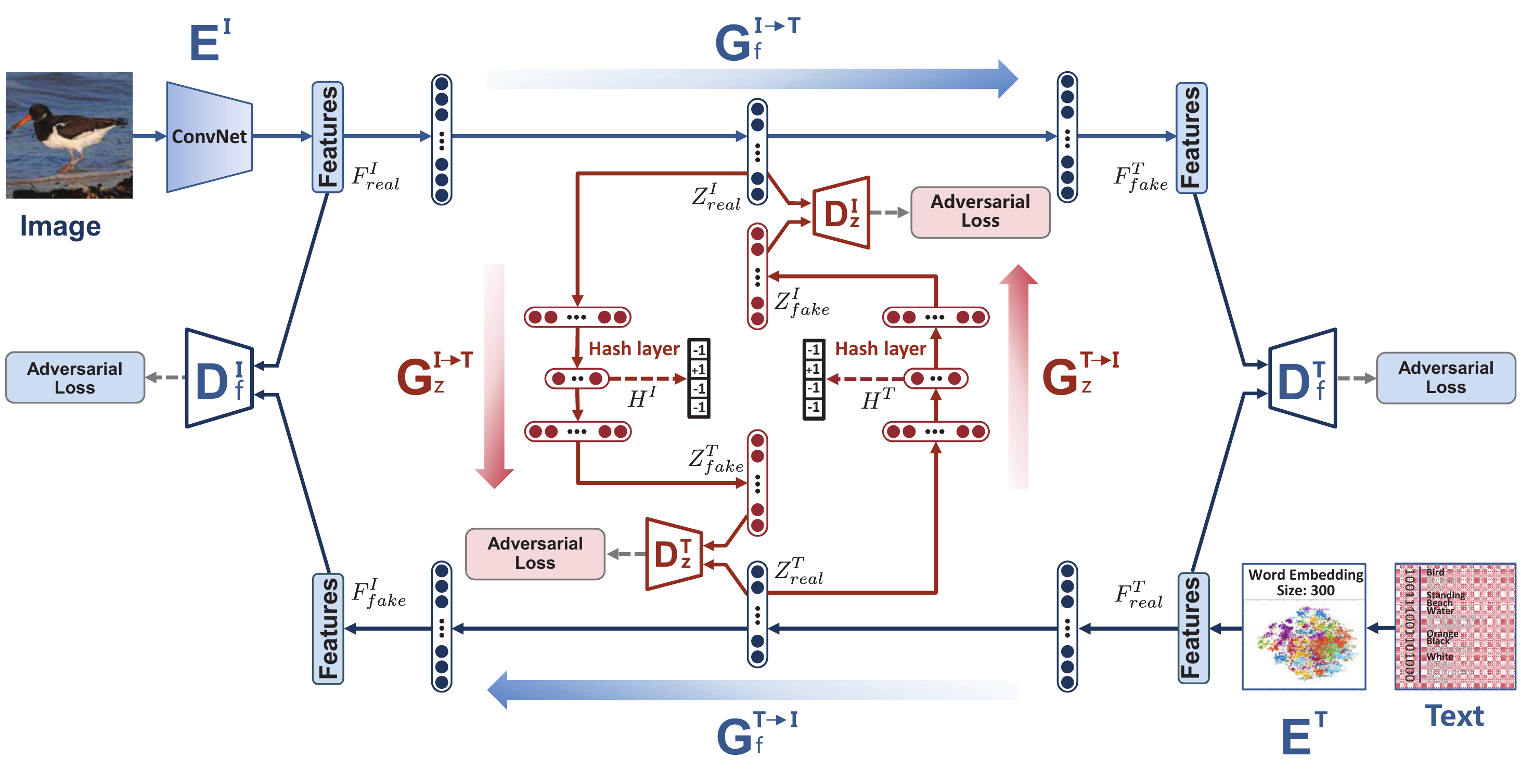}
	\end{center}	
	\caption{The proposed unsupervised coupled cycle generative adversarial deep cross-modal hashing learning framework. The main framework mainly consists of two modules: representation learning~(in color blue) and hashing learning~(in color red). The representation learning is constructed with $E^{I}$, $E^{T}$, $G^{I\rightarrow T}_{f}$, $G^{T\rightarrow I}_{f}$, $D^{I}_{f}$, and $D^{T}_{f}$, where $G^{I\rightarrow T}_{f}$, $D^{T}_{f}$, $G^{T\rightarrow I}_{f}$, and $D^{I}_{f}$ format the outer-cycle GAN. $G^{I\rightarrow T}_{z}$, $D^{T}_{z}$, $G^{T\rightarrow I}_{z}$, and $D^{I}_{z}$, formating the inner-cycle GAN, are combined into the hashing learning.}
	\label{fig::framework}
\end{figure*}
\section{Introduction}
Multi-modal data, such as text and image, exhibit heterogeneous properties, making it difficult for users to search for information of interest effectively and efficiently. Cross-modal retrieval, which aims to search the images (resp. texts) that are relevant to a given text (resp. image) query, has been studied in the past decade~\cite{Zhang2016CVPR,yang2017latent,gu2018look}. However, most of these methods suffer from high computation burden because of large volumes and high dimensions of multimedia data. Hashing based cross-modal retrieval, mapping multi-modal data into compact binary codes and conducting retrieval via fast bit-wise XOR operation, has become a hot topic. The fundamental challenges in cross-modal hashing retrieval lie in capturing the correlation between different modalities and learning reliable hash codes. 

Cross-modal hashing methods can be generally categorized into two groups: supervised and unsupervised. Supervised hashing methods~\cite{Liu2017CVPR,deng2016discriminative,Wu2014Sparse,Masci2014Multimodal,yang2017pairwise,li2018self} learn hash codes under some supervision such as label information. However, label information collection consumes massive time and labor, making it infeasible in real-world applications. Different from supervised scenario, unsupervised hashing methods~\cite{Bronstein2010Data,Kumar2011Learning,Feng2014Cross,yu2017binary,yang2018semantic} without using semantic labels, usually depend on the criterion of maintaining the original relationship among the training data. 

Recently, deep cross-modal hashing methods, leveraging the superior capacity of deep neural network~\cite{krizhevsky2012imagenet,goodfellow2014generative,zhu2017unpaired} to capture the correlation between different modalities, have illustrated that deep cross-modal hashing methods are usually more effective than shallow structure based counterparts. Deep Cross-Modal Hashing (DCMH)~\cite{Jiang2017CVPR}, Triplet based Deep Hashing (TDH)~\cite{deng2018triplet}, Shared Predictive Deep Quantization (SPDQ)~\cite{yang2018shared}, and Self-Supervised Adversarial Hashing (SSAH)~\cite{li2018self} are reported recently to encode individual modalities into their corresponding features by constructing two different pathways in deep networks. SPDQ constructs two specific network layers to learn modality-common and modality-private representations. DCMH and SSAH learn hash codes by preserving semantic correlation between different modalities which is constructed by label information. However, almost all of these supervised methods require amounts of label information. Compared with them, unsupervised deep hashing methods learn the modality correlation depending on correspondences between pairs of data (such as image-text pairs), making it more feasible in real word settings. Among these unsupervised deep hashing methods, Unsupervised Generative Adversarial Cross-modal Hashing (UGACH)~\cite{zhang2017unsupervised} exploits graph manifold structure to learn modality correlation. Unsupervised Deep Cross-Modal Hashing (UDCMH)~\cite{wu2018unsupervised} equips with adaptive learning strategy to learn hash codes iteratively. Even so, there are still some common disadvantages hindering these methods. First, due to plentiful semantically-similar pairs existing in real datasets, these methods just simply adopt individual network to extract features for each modality, which cannot build accurate correlation between different modalities. Second, most of these methods, separating hash codes generation from common representation learning, greatly decreases accuracy of the learned hash codes.

In this paper, we propose a novel unsupervised coupled cycle generative adversarial hashing network called UCH, for cross-modal retrieval. Specifically, we devise pair-coupled generative adversarial networks to build two cycle networks in an unified framework, where outer-cycle network is constructed to learn powerful representations for individual modality data to capture accurate correlation between different modalities and inner-cycle network is designed to generate reliable compact hash codes depending on the learned representations. The highlights of our work can be summarized as follows: 
\begin{itemize}
	\item We design an unsupervised coupled cycle generative adversarial hashing network for cross-modal retrieval, where powerful common representation and reliable hash codes can be learned simultaneously in an unified framework.
	\item By utilizing the proposed coupled cycle networks, common representation and hash codes learning can interact with each other and reach the optimum when network is convergence in the meanwhile.
	\item Extensive experiments conducted on three popular benchmark datasets demonstrate that our proposed UCH outperforms the state-of-the-art unsupervised cross-modal hashing methods.  
\end{itemize}  

The rest of this paper starts with a review of the most related work. Then the proposed unsupervised coupled cycle generative adversarial deep cross-modal hashing method is presented in detail, which is followed by the experiments and conclusion.

\section{Related Work}
In this section, the most related work on topic of unsupervised cross-modal hashing methods are reviewed, both traditional shallow structure based methods and deep structure based methods.

For shallow structure based unsupervised cross-modal hashing methods, co-occurrence information is uniformly utilized to exploit modality correlation since image-text pair that occurs simultaneously tend to be of similar semantic.  In~\cite{rasiwasia2010new}, canonical component analysis (CCA) is adopted to map image-text pairs into a latent space where their similarities can be measured. Cross-view hashing (CVH)~\cite{Kumar2011Learning} considers the cross-modality similarity preservation during hashing learning process. In~\cite{rastegari2013predictable}, Predictable Dual-View Hashing (PDH) learns hashing function in virtue of self-taught learning algorithm. In Collective Matrix Factorization Hashing (CMFH)~\cite{Ding2014CVPR}, unified hash codes are learned by collective matrix factorization from different views. Latent Semantic Sparse Hashing (LSSH)~\cite{zhou2014latent} is proposed to jointly learn latent features from images and texts with sparse coding. In Fusion Similarity Hashing (FSH)~\cite{Liu2017CVPR}, modality correlation is captured by fusing similarity across different modalities.

For deep structure based unsupervised cross-modal hashing methods, deep networks pretrained on other public dataset are usually utilized to predict individual modality representation and hash codes further can be learned based on the achieved representation. UGACH presents that a similarity relationship matrix depending on its extracted representation is first constructed and then modality correlation is learned by exploiting manifold structure between different modalities. However, there are two problems indwelling in UGACH. One is that an accurate similarity matrix cannot be obtained through only one calculation, the other is that time consuming in calculating similarity matrix cannot be ignored. UDCMH is another representative unsupervised deep hashing method, which applies matrix factorization on the extracted deep features to generate hash codes and then uses the obtained hash codes as supervision to update feature extractor network. However, hash codes usually have short bit representations, whose supervised information is insufficient to learn discriminative representations for exploiting the modality correlation. In contrast, our UCH can effectively build modality correlation by directly learning bidirectional transformation between different modalities. Additionally, with the proposed coupled cycle framework, hash codes can be updated iteratively, which are more reliable.

\section{Proposed UCH}
Fig.~\ref{fig::framework} shows the overview of our proposed UCH for cross-modal retrieval, which mainly consists of two modules: representation learning~(in color blue) and hashing learning~(in color red). Specifically, the representation learning part consists of the $E^{I}$, $E^{T}$, $G^{I\rightarrow T}_{f}$, $G^{T\rightarrow I}_{f}$, $D^{I}_{f}$, and $D^{T}_{f}$, where $G^{I\rightarrow T}_{f}$, $D^{T}_{f}$, $G^{T\rightarrow I}_{f}$, and $D^{I}_{f}$ form a  generative adversarial cycle network, namely outer-cycle GAN. By using $E^{I}$ and $E^{T}$, real image features $F^{I}_{real}$ and real text features $F^{T}_{real}$ are extracted from original image and text. To learn powerful cross-modal common representations $Z^{I}_{real}$ and $Z^{T}_{real}$, each modality data is first mapped into a shared common space, and then to generate different modality data. Additionally, hashing learning part consisting of $G^{I\rightarrow T}_{z}$, $D^{T}_{z}$, $G^{T\rightarrow I}_{z}$, and $D^{I}_{z}$ are combined to form a generative adversarial cycle network, namely inner-cycle GAN. To further build the projection from original image and text to hash codes, we feed the learned $Z^{I}_{real}$ and $Z^{T}_{real}$ into $G^{I\rightarrow T}_{z}$ and $G^{T\rightarrow I}_{z}$ to generate $Z^{T}_{fake}$ and $Z^{I}_{fake}$ recorded as fake image and fake text representation. Two hashing layers are devised within the inner-cycle GAN to generate hash codes directly. The inner-cycle GAN and outer-cycle GAN are trained jointly. Overall, through the outer-cycle GAN network of cross-modal feature learning, we hope to learn powerful cross-modal common representations, and through the inner-cycle GAN network of hashing learning, we hope to learn reliable hash codes. In the following, we first give some definitions and then introduce the proposed method in detail.
\subsection{Problem Definition}
Let $O=\{o_{i}\}_{i=1}^{n}$ denote one cross-modal dataset, $o_{i}=(v_{i}, t_{i})$ is an image-text pair, where $v_{i}$ and $t_{i}$ are raw visual and textual information describing the $i$th instance. The goal of cross-modal hashing is to generate reliable hash codes for image and text: $B^{\ast} \in \{-1,1\}^{n\times K}$, $\ast =\{v,t\}$, where $K$ is the length of binary codes. In our UCH, the outputs of hashing layers are defined as $H^{\ast}$, $\ast\in\{v,t\}$ for image and text respectively, binary hash codes $B^{\ast}$ are generated by applying a sign function to $H^{\ast}$:
\begin{equation} \label{eq::hashing}
B^{\ast} \ = \ sign(H^{\ast}), \ast \in \{v,t\},
\end{equation}

\subsection{Representation Learning}
Our goal is to obtain reliable hash codes for image and text. However, in view of the huge modality gap between two modalities caused by their heterogeneous structures, modality correlation cannot be learned directly without using any supervision information. To build the relationship between different modalities, we construct outer-cycle GAN with two generator networks for image and text, making it possible to generate image~(resp. text) modality data from text~(resp. image) modality data. By training these two networks, powerful individual representation can be learned for different modalities and thus modality gap can be bridged effectively. Given image-text pair: $v$ and $t$, $F^{I}_{real}$ and $F^{T}_{real}$ are extracted through $E^{I}(v, \theta^{I})$ and $E^{t}(t, \theta^{T})$, where $\theta^{I}$ and $\theta^{T}$ are network parameters. Within the outer-cycle GAN, the generation can be formulated as:
\begin{equation} \label{eq::feature gen}
	\begin{aligned}
	&F^{T}_{fake} = G^{I\rightarrow T}_{f}(F^{I}_{real},\eta^{I\rightarrow T}), \\
	&F^{I}_{fake} = G^{T\rightarrow I}_{f}(F^{T}_{real},\eta^{T\rightarrow I}),
	\end{aligned}
\end{equation} 
where $G^{I\rightarrow T}_{f}$ and $G^{T\rightarrow I}_{f}$ are two generation functions, $\eta^{I\rightarrow T}$ and $\eta^{T\rightarrow I}$ are network parameters. 

Moreover, to judge the quality of the generative data, we further devise two discriminators $D^{I}_{f}(\cdot, \mu^{I})$ and $D^{T}_{f}(\cdot, \mu^{T})$ to provide adversarial loss for generators, where $\mu^{I}$ and $\mu^{T}$ are network parameters. For different modalities image and text, their individual adversarial losses $\mathcal{L}_{adv\_f}$ can be written as:
\begin{equation} \label{eq::Adver Outer}
	\mathcal{L}_{adv\_f} = \mathcal{L}^{I}_{adv\_f} + \mathcal{L}^{T}_{adv\_f},
\end{equation}
where $\mathcal{L}^{I}_{adv\_f}$ and $\mathcal{L}^{T}_{adv\_f}$ are formulated as follows:
\begin{equation} \label{eq::Adver Outer Image}
	\begin{aligned}
	\mathcal{L}^{I}_{adv\_f} = &\min_{\eta ^{T\rightarrow I}} \max_{\mu^{I}} \ E_{F^{I}_{real}\sim P(F^{I}_{real})}\left[ \log D^{I}_{f}(F^{I}_{real})\right] \\
	+ &E_{F^{I}_{fake}\sim P(F^{I}_{fake})}\left[\log(1-D^{I}_{f}(F^{I}_{fake})) \right],
	\end{aligned}
\end{equation}
\begin{equation} \label{eq::Adver Outer Text}
	\begin{aligned}
	\mathcal{L}^{T}_{adv\_f} = &\min_{\eta ^{I\rightarrow T}} \max_{\mu^{T}} \ E_{F^{T}_{real}\sim P(F^{T}_{real})}\left[ \log D^{T}_{f}(F^{T}_{real})\right] \\
	+ & E_{F^{T}_{fake} \sim P(F^{T}_{fake})}\left[\log(1-D^{T}_{f}(F^{T}_{fake})) \right].
	\end{aligned}
\end{equation}

Furthermore, to improve the generative capacity of the outer-cycle GAN networks, we feed the generated $F^{T}_{fake}$ and $F^{I}_{fake}$ back into the generative networks to reconstruct the original data and minimize the reconstruction loss, which can be formulated as:
\begin{equation} \label{eq::recon loss }
	\begin{aligned}
	\mathcal{L}_{rec\_f} =& \min \ \mathcal{L}^{I}_{rec\_f} + \mathcal{L}^{T}_{rec\_f} \\
	= \min \sum_{i=1}^{n}& E_{F^{I}_{real}\sim P(F^{I}_{real})}\left[\lVert F^{I}_{real}-G^{T\rightarrow I}(F^{T}_{fake})\rVert_2^2\right] \\
	+ \sum_{i=1}^{n}&E_{F^{T}_{real}\sim P(F^{T}_{real})}\left[\lVert F^{T}_{real}-G^{I\rightarrow T}(F^{I}_{fake})\rVert_2^2\right].
	\end{aligned}
\end{equation}

Finally, we hope to generate similar representations $Z^{I}_{real}$ and $Z^{T}_{real}$ for semantically-similarity paired cross-modal data. $l_{2}$ loss function is adopted in the shared common space to regulate the learned representation to be similar. Thus, the similarity loss can be achieved by:
\begin{equation} \label{eq::sim loss}
	\begin{aligned}
	\mathcal{L}_{sim\_f} = \min \sum_{i=1}^{n}\lVert Z^{I}_{real}-Z^{T}_{real}\rVert_2^2.
	\end{aligned}
\end{equation}

As to the representation learning, the whole loss function to learn powerful individual representation for different modalities is made up of adversarial loss, reconstruction loss, and similarity loss, which can be defined as:
\begin{equation}\label{eq::repre loss}
	\mathcal{L}_{f} = \mathcal{L}_{adv\_f} + \mathcal{L}_{rec\_f} + \mathcal{L}_{sim\_f}.
\end{equation}

Through training the outer-cycle GAN to minimize (\ref{eq::repre loss}), discriminative representations can be achieved. 

\subsection{Hashing Learning}
With the similar idea that different modality data can be inherited by the common representations in representation learning part, we learn hash codes within the inner-cycle GAN constructed by $G^{I\rightarrow T}_{z}(\cdot,\xi^{I\rightarrow T})$, $D^{T}_{z}(\cdot, \varepsilon^{T})$, $G^{T\rightarrow I}_{z}(\cdot, \xi^{T\rightarrow I})$, and $D^{I}_{z}(\cdot, \varepsilon^{I})$, where $\xi^{I\rightarrow T}$, $\varepsilon^{T}$, $\xi^{T\rightarrow I}$, and $\varepsilon^{I}$ are network parameters. Similarly, given $Z^{I}_{real}$ and $Z^{T}_{real}$, the generation can be formulated as:
\begin{equation} \label{eq::hash gen}
	\begin{aligned}
	&Z^{T}_{fake} = G^{I\rightarrow T}_{z}(Z^{I}_{real},\xi^{I\rightarrow T}) \\
	&Z^{I}_{fake} = G^{T\rightarrow I}_{z}(Z^{T}_{real},\xi^{T\rightarrow I}).
	\end{aligned}
\end{equation} 

Further, the adversarial loss for hashing learning $\mathcal{L}_{adv\_z} = \mathcal{L}^{I}_{adv\_z} + \mathcal{L}^{T}_{adv\_z}$, where $\mathcal{L}^{I}_{adv\_z}$ and $\mathcal{L}^{T}_{adv\_z}$ are designed as :
\begin{equation} \label{eq::Adver Inter Image}
	\begin{aligned}
	\mathcal{L}^{I}_{adv\_z} = \min_{\xi ^{T\rightarrow I}} \max_{\varepsilon^{I}} \ E_{Z^{I}_{real}\sim P(Z^{I}_{real})}\left[ \log D^{I}_{z}(Z^{I}_{real})\right]& \\
	+ E_{Z^{I}_{fake}\sim P(Z^{I}_{fake})}\left[\log(1-D^{I}_{z}(Z^{I}_{fake})) \right]&.
	\end{aligned}
\end{equation}
\begin{equation} \label{eq::Adver Inter Text}
	\begin{aligned}
	\mathcal{L}^{T}_{adv\_z} = \min_{\xi ^{I\rightarrow T}} \max_{\varepsilon^{T}} \ E_{Z^{T}_{real}\sim P(Z^{T}_{real})}\left[ \log D^{T}(Z^{T}_{real})\right]& \\
	+ E_{Z^{T}_{fake} \sim P(Z^{T}_{fake})}\left[\log(1-D^{T}(Z^{T}_{fake})) \right]&.
	\end{aligned}
\end{equation}

Next, reconstruction loss $\mathcal{L}_{rec\_z}$ is constructed as:
\begin{equation} \label{eq::recon loss hash}
	\begin{aligned}
	\mathcal{L}_{rec\_z} =& \min \ \mathcal{L}^{I}_{rec\_z} + \mathcal{L}^{t}_{rec\_z} \\
	= \min \sum_{i=1}^{n}& E_{Z^{I}_{real}\sim P(Z^{I}_{real})}\left[\lVert Z^{I}_{real}-G^{T\rightarrow I}(Z^{T}_{fake})\rVert_2^2\right] \\
	+ \sum_{i=1}^{n}&E_{Z^{T}_{real}\sim P(Z^{T}_{real})}\left[\lVert Z^{T}_{real}-G^{I\rightarrow T}(Z^{I}_{fake})\rVert_2^2\right].
	\end{aligned}
\end{equation}

Finally, similarity loss $\mathcal{L}_{sim\_z}$ is utilized to promote our UCH to produce uniformed hash codes, which can be written as:
\begin{equation} \label{eq::sim loss hash}
	\begin{aligned}
	\mathcal{L}_{sim\_z} = \min \sum_{i=1}^{n}\lVert H^{I}-H^{T}\rVert_2^2.
	\end{aligned}
\end{equation}

Therefore, the final loss for hashing learning within inner-cycle GAN can be written as:
\begin{equation}\label{eq::hash loss}
	\mathcal{L}_{z} = \mathcal{L}_{adv\_z} + \mathcal{L}_{rec\_z} + \mathcal{L}_{sim\_z}.
\end{equation}

Overall, the whole loss for the entire framework mainly consists of two parts: representation learning loss and hashing learning loss, which is written as:
\begin{equation}\label{eq::total loss}
	\mathcal{L}_{Total} = \mathcal{L}_{f} + \mathcal{L}_{z}.
\end{equation}

The whole alternating learning algorithm for the proposed UCH is briefly outlined in Algorithm~\ref{alg::algorithm}.
	\begin{algorithm}[!t]
		\algsetup{}
		\caption{Optimizing process of the proposed UCH}
		\label{alg::algorithm}
		\begin{algorithmic}
			\REQUIRE Image-Text pairs dataset;
			\ENSURE Optimal code matrix $B$
			\STATE \textbf{Initialization} \STATE Initialize network parameters $\theta^{\ast}$, $\mu^{\ast}$, $\varepsilon^{\ast}$, $\eta^{I\rightarrow T}$, $\eta^{T\rightarrow I}$, $\xi^{I\rightarrow T}$, and $\xi^{T\rightarrow I}$, where $\ast=\{v,t\}$, batch size: $N^{I,T}=128$, maximum iteration number: $T_{max}$.
			\REPEAT
			\STATE Update $\mu^{I}$ and $\mu^{T}$ by~\eqref{eq::opt_repre dis} with BP algorithm;
			\STATE Update $\theta^{I}$, $\theta^{T}$, $\eta^{I\rightarrow T}$, and $\eta^{T\rightarrow I}$ by~\eqref{eq::opt_repre gen} with BP algorithm;
			\STATE Update $\varepsilon^{I}$ and $\varepsilon^{I}$ by~\eqref{eq::opt_hash dis} with BP algorithm;
			\STATE Update $\xi^{I\rightarrow T}$ and $\xi^{T\rightarrow I}$ by~\eqref{eq::opt_hash gen} with BP algorithm;
			\STATE Update hash codes matrix $B$ by $B=sign(H^{I}+H^{T})$;
			\UNTIL{maximum iteration number $T_{max}$}.
		\end{algorithmic}
	\end{algorithm}
\subsection{Optimization}
Considering the gradient vanishing problem caused by the minimax loss for generative adversarial networks, we optimize our UCH separately. 

In order to guarantee the learning efficiency of the proposed UCH, the outer-cycle GAN is trained to produce powerful representations firstly. Update $\mu^{I}$ and $\mu^{T}$, with the other parameters fixed:
\begin{equation} \label{eq::opt_repre dis}
{\mu^{I}, \mu^{T}} = \arg \max_{ ^{\mu^{I}, \mu^{T}}} \mathcal{L}_{adv\_f}.
\end{equation}

Update $\theta^{I}$, $\theta^{T}$, $\eta^{I\rightarrow T}$, and $\eta^{T\rightarrow I}$ by minimizing (\ref{eq::repre loss}) with the other parameters fixed:
\begin{equation} \label{eq::opt_repre gen}
\begin{aligned}
{\theta^{I}, \theta^{T}, \eta^{I\rightarrow T}, \eta^{T\rightarrow I}} = \arg\min_{\underset {  \eta^{I\rightarrow T}, \eta^{T\rightarrow I}}{ \theta^{I}, \theta^{T} }  } \mathcal{L}_{rec\_f} + \mathcal{L}_{sim\_f}. 
\end{aligned}
\end{equation}
\begin{table*}[!t]
	\begin{center}
		\caption{Comparison in terms of MAP scores of two retrieval tasks on MIRFlickr-25K, IAPR TC-12, and COCO datasets with different lengths of hash codes. The best accuracy are shown in boldface.}
		\label{result::MAP}
		\begin{tabular}{|c|c|c|c|c|c|c|c|c|c|c|}
			\hline
			\multirow{2}{*}{Task} &\multirow{2}{*}{Method} &\multicolumn{3}{c|}{MIRFlickr-25K}&\multicolumn{3}{c|}{IAPR TC-12}&\multicolumn{3}{c|}{COCO}\\
			\cline{3-11}
			& & 16 & 32 & 64 & 16 & 32 & 64 & 16 & 32 & 64 \\
			\hline
			\multirow{8}{*}{\tabincell{c}{Image Query \\ v.s.\\Text Database}}
			& CVH         & 0.596 & 0.585 & 0.582  & 0.397 & 0.385 & 0.377 & 0.474 & 0.475 & 0.459 \\
			& STMH        & 0.575 & 0.582 & 0.625  & 0.374 & 0.387 & 0.400 & 0.405 & 0.414 & 0.406 \\
			& LSSH        & 0.589 & 0.604 & 0.634  & 0.443 & 0.456 & 0.460 & 0.468 & 0.473 & 0.485 \\
			& FSH         & 0.580 & 0.583 & 0.591  & 0.400 & 0.420 & 0.422 & 0.453 & 0.489 & 0.494 \\	
			& CMFH        & 0.601 & 0.605 & 0.610  & 0.442 & 0.450 & 0.463 & 0.475 & 0.490 & 0.524 \\
			\cline{2-11}
			& CMSSH       & 0.598 & 0.593 & 0.600  & 0.390 & 0.386 & 0.377 & 0.503 & 0.516 & 0.517 \\
			& SCM         & 0.635 & 0.636 & 0.639  & 0.402 & 0.412 & 0.419 & 0.442 & 0.461 & 0.486 \\
			\cline{2-11}
			& \textbf{OURS}  & $\mathbf{0.654}$ & $\mathbf{0.669}$ & $\mathbf{0.679}$  & $\mathbf{0.447}$ & $\mathbf{0.471}$ & $\mathbf{0.485}$ & $\mathbf{0.521}$ & $\mathbf{0.534 }$ & $\mathbf{0.547}$ \\
			\hline
			\multirow{8}{*}{\tabincell{c}{Text Query \\  v.s.\\Image Database}}
			& CVH         & 0.604 & 0.592 & 0.577 & 0.405 & 0.393 & 0.384 & 0.470 & 0.474 & 0.456\\
			& STMH        & 0.586 & 0.594 & 0.632 & 0.381 & 0.406 & 0.429 & 0.391 & 0.422 & 0.449\\
			& LSSH        & 0.583 & 0.588 & 0.601 & 0.409 & 0.416 & 0.421 & 0.456 & 0.460 & 0.463\\
			& FSH         & 0.575 & 0.576 & 0.583 & 0.412 & 0.429 & 0.444 & 0.471 & 0.509 & 0.514\\	
			& CMFH        & 0.627 & 0.636 & 0.639 & 0.435 & 0.445 & 0.456 & 0.468 & 0.478 & 0.509\\
			\cline{2-11}
			& CMSSH       & 0.563 & 0.571 & 0.566 & 0.385 & 0.383 & 0.393 & 0.431 & 0.434 & 0.469\\
			& SCM         & 0.652 & 0.657 & 0.659 & 0.437 & 0.450 & 0.458  & 0.426 & 0.442 & 0.461\\
			\cline{2-11}
			& \textbf{OURS}  & $\mathbf{0.661}$ & $\mathbf{0.667}$ & $\mathbf{0.668}$  & $\mathbf{0.446}$ & $\mathbf{0.469}$ & $\mathbf{0.488}$ & $\mathbf{0.499 }$ & $\mathbf{0.519 }$ & $\mathbf{0.545}$\\
			\hline
		\end{tabular}
	\end{center}
\end{table*}

Then, train the inner-cycle GAN to generate reliable hash codes. Update $\varepsilon^{I}$ and $\varepsilon^{T}$ with other parameters fixed:
\begin{equation} \label{eq::opt_hash dis}
{\varepsilon^{I}, \varepsilon^{T}} = \arg \max_{\varepsilon^{I}, \varepsilon^{T}} \mathcal{L}_{adv\_z}.
\end{equation}

Finally, update $\xi^{I\rightarrow T}$ and $\xi^{T\rightarrow I}$ with other parameters fixed:
\begin{equation} \label{eq::opt_hash gen}
{\xi^{I\rightarrow T}, \xi^{T\rightarrow I}} = \arg \min_{\xi^{I\rightarrow T}, \xi^{T\rightarrow I}}  \mathcal{L}_{rec\_z} + \mathcal{L}_{sim\_z}. 
\end{equation}

\section{Experiments}
\subsection{Datasets}
Three popular benchmark datasets in cross-modal retrieval:~\emph{MIRFlickr-25K}~\cite{huiskes2008mir},~\emph{NUS-WIDE}~\cite{chua2009nus}, and~\emph{Microsoft COCO}~\cite{lin2014microsoft} are adopted to validate our proposed method. 

\emph{MIRFlickr-25K} dataset consists of 25,000 images collected from Flickr website. Each image is associated with multiple textual tags and manually annotated with at least one of the 24 unique labels. Leaving out data without labeled information, 20,015 image-text pairs are used in our experiment totally, where 2,000 image-text pairs are randomly selected as query set and the rest are regarded as retrieval set. The textual information for each image is represented as a 1386-dimensional bag-of-words vector. For supervised baselines, 5,000 image-text pairs are selected from retrieval set to construct training set. 

\emph{IAPR TC-12} dataset contains 20,000 images with corresponding sentence descriptions. These image-text pairs present various semantics such as action and people categories. In our experiment, each sentence is represented as a bag-of-words vector with 2,000-dimension. We prune all images without any labeled information and finally 18,571 image-text pairs are remained. Each image-text pair belongs to the top 22 frequent labels from the 275 concepts. We randomly select 5,000 image-text pairs to construct training set for supervised methods and the rest are retrieval set. 

\emph{Microsoft COCO} dataset totally contains 82,783 training images and 40,504 validation images. Each image is described with five different sentences and labeled with at least one of 80 unique labels. In our experiment, we represent each sentence with a 2,000-dimension bag-of-words vector. After deleting the image-text pairs without any textual information, finally, 122,218 image-text pairs are left to formulate the dataset used in our experiment. 2,000 image-text pairs are randomly selected as query set and the left 120,218 pairs are regarded as retrieval set. 6,000 image-text pairs from retrieval set are randomly selected to construct training set for supervised methods. Additionally, it should be noted that all retrieval set are used as training set for all unsupervised methods.
\label{Experiment Results}
\begin{figure*}
	\centering
	\subfloat[]{
		\begin{minipage}[b]{.33\linewidth}
			\centering
			\includegraphics[width=.99\textwidth]{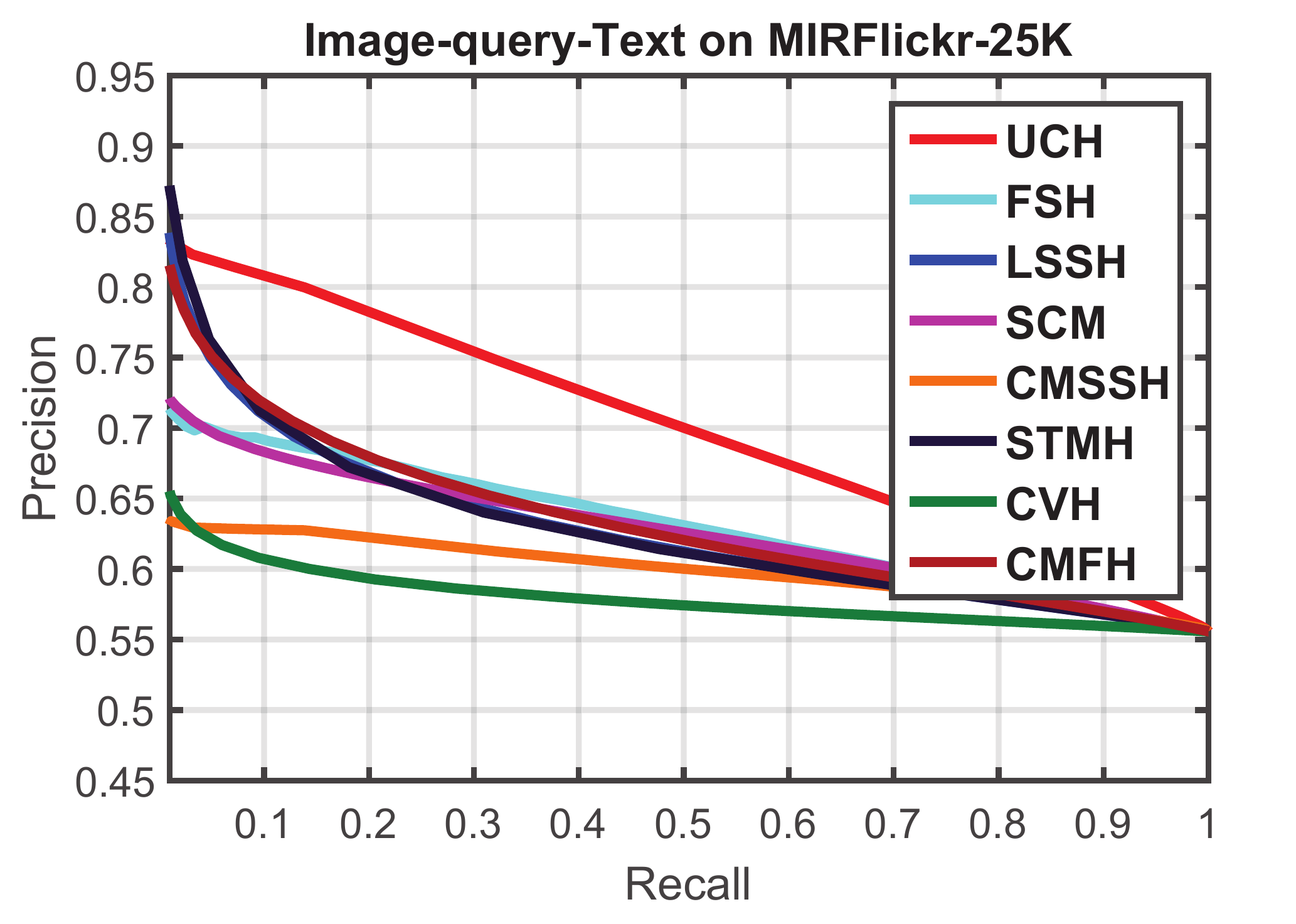}
	\end{minipage}}
	\subfloat[]{
		\begin{minipage}[b]{.33\linewidth}
			\centering
			\includegraphics[width=.99\textwidth]{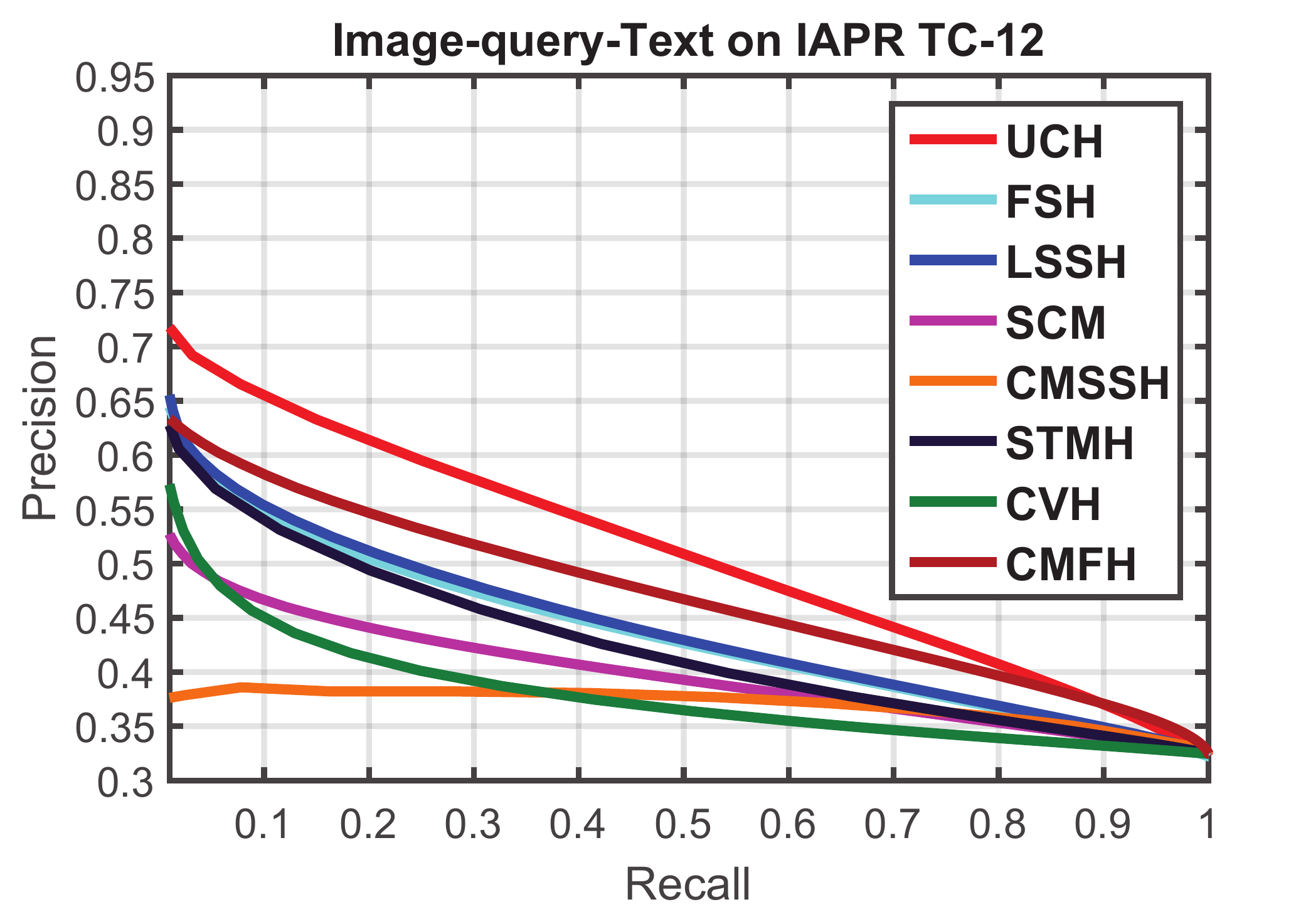}
	\end{minipage}}
	\subfloat[]{
		\begin{minipage}[b]{.33\linewidth}
			\centering
			\includegraphics[width=.99\textwidth]{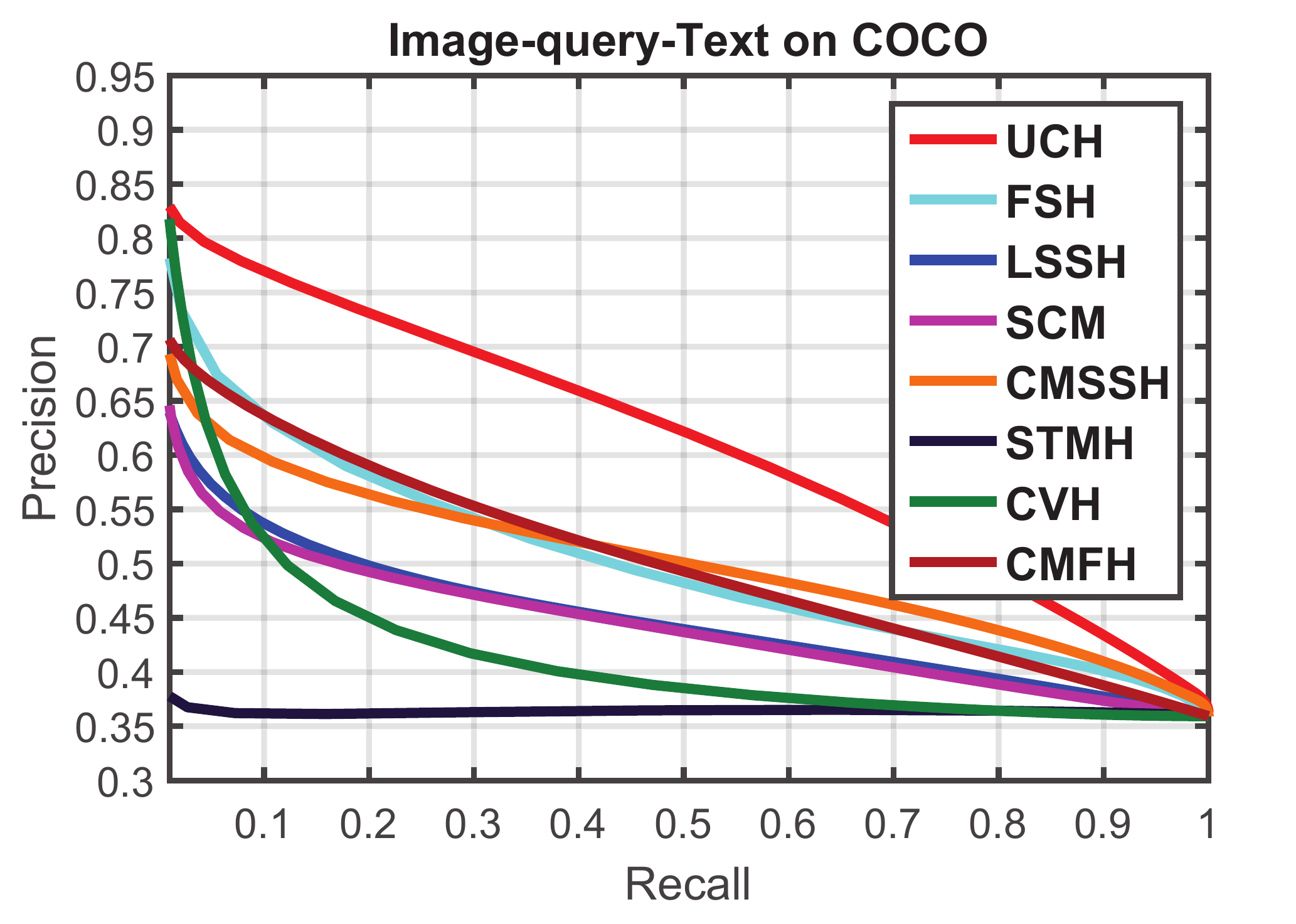}
	\end{minipage}}\\
	\subfloat[]{
		\begin{minipage}[b]{.33\linewidth}
			\centering
			\includegraphics[width=.99\textwidth]{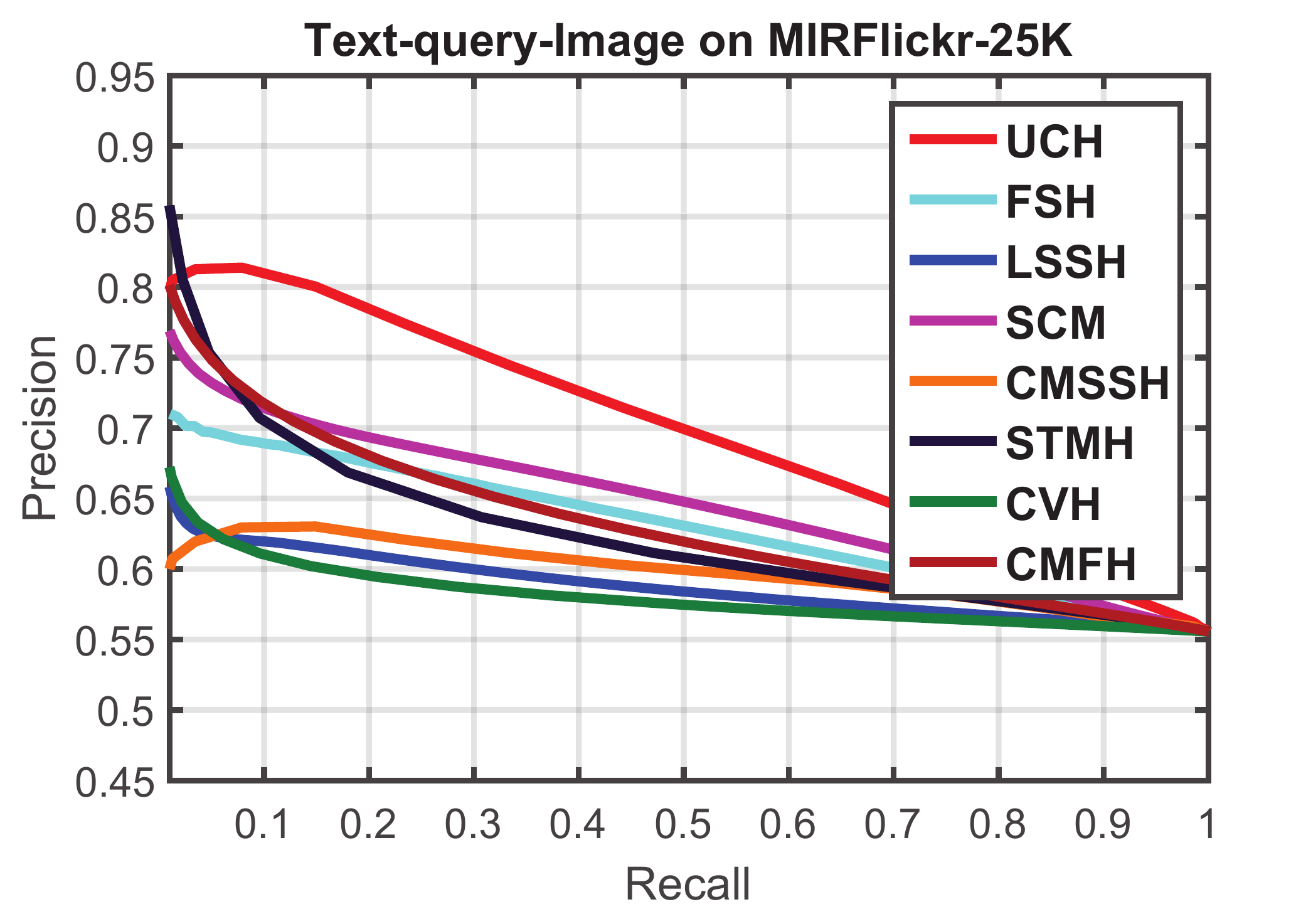}
	\end{minipage}}
	\subfloat[]{
		\begin{minipage}[b]{.33\linewidth}
			\centering
			\includegraphics[width=.99\textwidth]{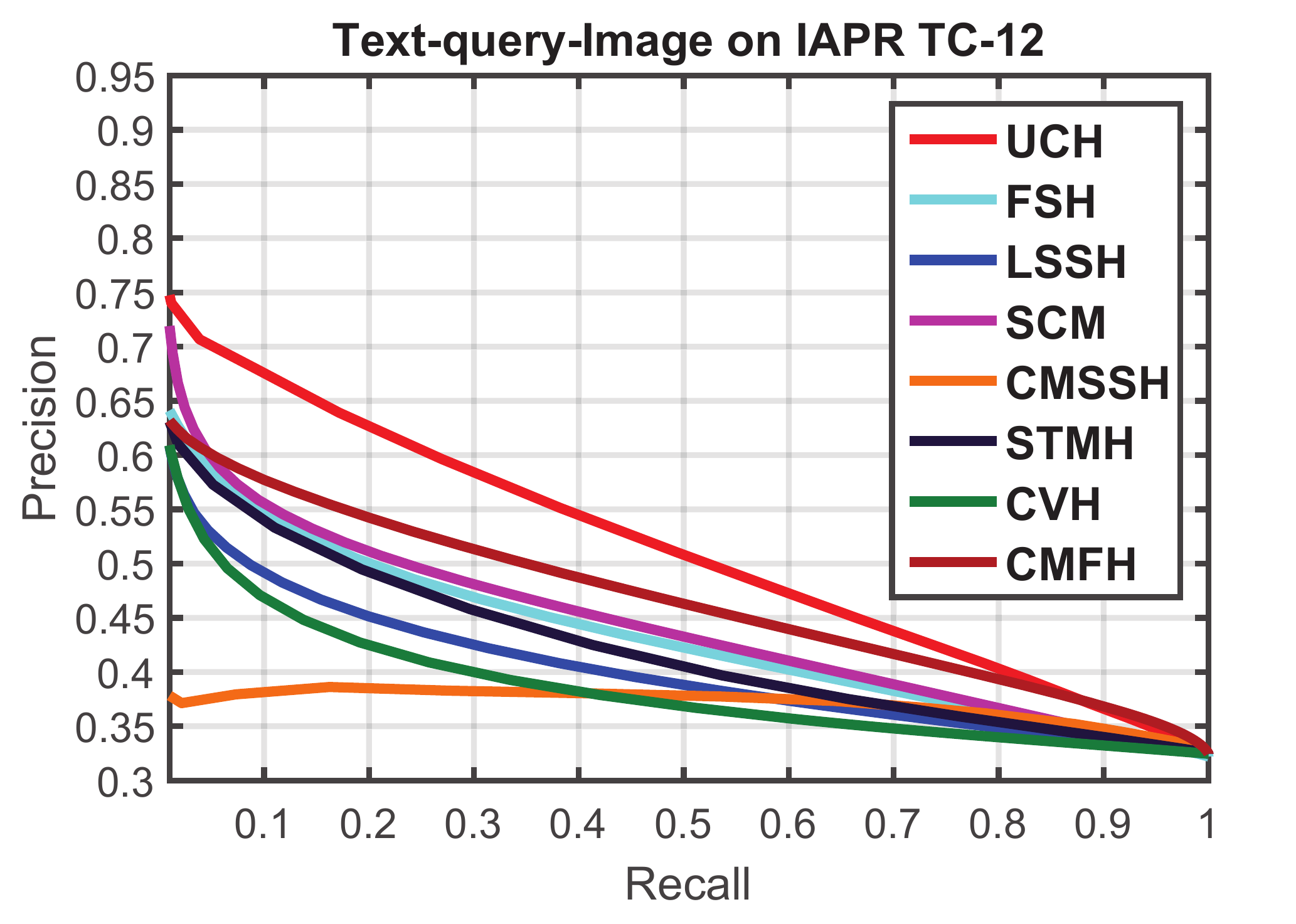}
	\end{minipage}}
	\subfloat[]{
		\begin{minipage}[b]{.33\linewidth}
			\centering
			\includegraphics[width=.99\textwidth]{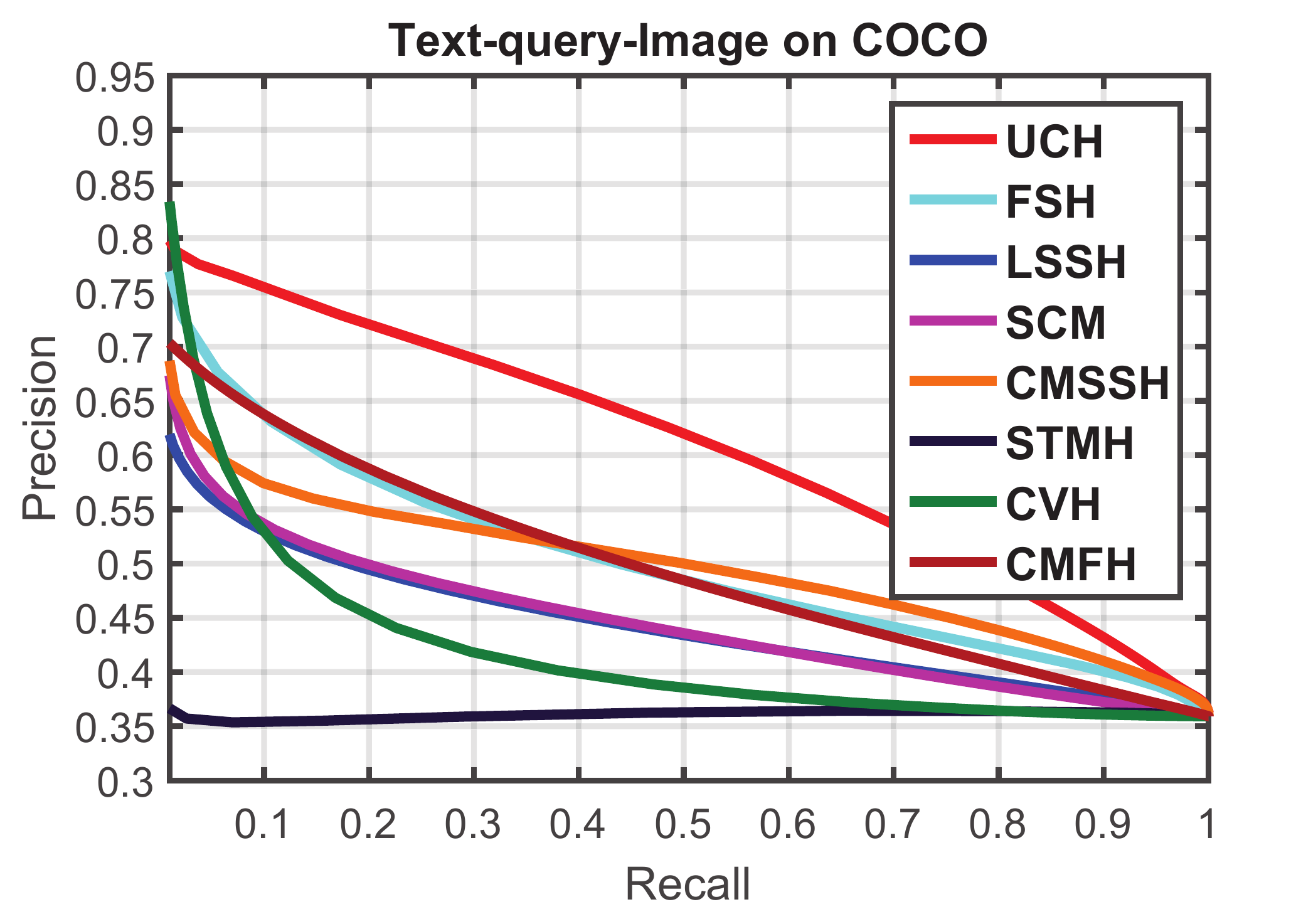}
	\end{minipage}}
	\caption{Precision-recall curves evaluated on MIRFlickr25k, IAPR TC-12, and COCO. The code length is 64.}
	\label{fig::PRcurve}
\end{figure*}

\subsection{Baselines and Evaluation}
To illustrate the effectiveness of our proposed UCH, there are seven state-of-the-art cross-modal hashing baselines compared in experiments, including five unsupervised methods CVH~\cite{Kumar2011Learning}, CMFH~\cite{Ding2014CVPR}, STMH~\cite{Wang2015Semantic}, LSSH~\cite{zhou2014latent}, and FSH~\cite{Liu2017CVPR} and two supervised methods CMSSH~\cite{Bronstein2010Data} and SCM~\cite{Zhang2014Large}, which are all shallow structure based cross-modal retrieval hashing methods. We additionally compare our UCH with the recently proposed UGACH~\cite{zhang2017unsupervised}, which is a deep structure based cross-modal retrieval hashing method. For fair comparison, we use the same deep network CNN-F to extract deep features for all shallow structure based methods. 

Following the previous methods, three common used protocols: Mean Average Precision~(MAP), Precision-Recall curves~(PR curves), and Precision of the top N curves~(Precision\emph{@}N), are adopted to evaluate the retrieval performance of all methods in our experiments.

\subsection{Implementation Details}
\label{section:Implementation Details}
In this subsection, we will introduce the implementation details about the proposed UCH and settings of some parameters. UCH is implemented via~\emph{TensorFlow} and executed on a server with two NVIDIA TITAN X GPUs.  

\textbf{Representation Learning:} $\mathbf{E^{I}}$, $\mathbf{E^{T}}$, $\mathbf{G^{I\rightarrow T}_f}$, $\mathbf{G^{T\rightarrow I}_f}$, $\mathbf{D^{I}_{f}}$, and $\mathbf{D^{T}_{f}}$. Our UCH is a CNN-based method, where CNN-F~\cite{chatfield2014return} neural network pretrained on ImageNet dataset~\cite{simonyan2014very} is adopted as image extractor. For image, the input raw image is resized into $224\times224\times3$ and feed into the CNN-F. We take the output of $fc7$ as real image features $F^{I}_{real}$. For text, we design an embedding network layer with 300 nodes following the text input layer to project raw bag-of-words vector into features with continuous values. We take the output of embedding layer as real text features $F^{I}_{real}$. $G^{I\rightarrow T}_f$ and $G^{T\rightarrow I}_f$ are constructed with two different deep networks with four full-connected layers,~\emph{e.g.,}~($G^{I\rightarrow T}_f$:$4096\rightarrow512\rightarrow256\rightarrow512\rightarrow300$ and $G^{T\rightarrow I}_f$: $300\rightarrow512\rightarrow256\rightarrow512\rightarrow4096$). $F^{I}_{real}$ and $F^{I}_{fake}$ are fed into $G^{I\rightarrow T}_f$ and $F^{T}_{real}$ and $F^{T}_{fake}$ are fed into $G^{T\rightarrow I}_f$. Meanwhile, two discriminator networks $D^{I}_{f}$ and $D^{T}_{f}$ are framed with two deep networks with two full-connected layers: $D^{I}_{f}$,~\emph{e.g.,}~($4096\rightarrow256\rightarrow32$) and $D^{T}_{f}$,~\emph{e.g.,}~($300\rightarrow256\rightarrow32$). 

\textbf{Hashing Learning: $\mathbf{G^{I\rightarrow T}_{z}}$, $\mathbf{G^{T\rightarrow I}_{z}}$, $\mathbf{D^{I}_{z}}$, and $\mathbf{D^{T}_{z}}$.} $G^{I\rightarrow T}_{z}$ and $G^{T\rightarrow I}_{z}$ are uniformly framed with four full-connected layers,~\emph{e.g.,}~($256\rightarrow128\rightarrow K\rightarrow128\rightarrow256$), $K$ is the queried code length. The discriminator networks $D^{I}_{z}$ and $D^{T}_{z}$ are framed with two full-connected layers,~\emph{e.g.,}~($256\rightarrow128\rightarrow32$). In all our experiments, the initial leaning rates of image and text networks are set to $10^{-4}$ and $10^{-2}$. And batchsize and weight decay are set to 128 and $10^{-1}$.
\begin{figure*}[!t]
	\centering
	\subfloat[]{
		\begin{minipage}[b]{.33\linewidth}
			\centering
			\includegraphics[width=.99\textwidth]{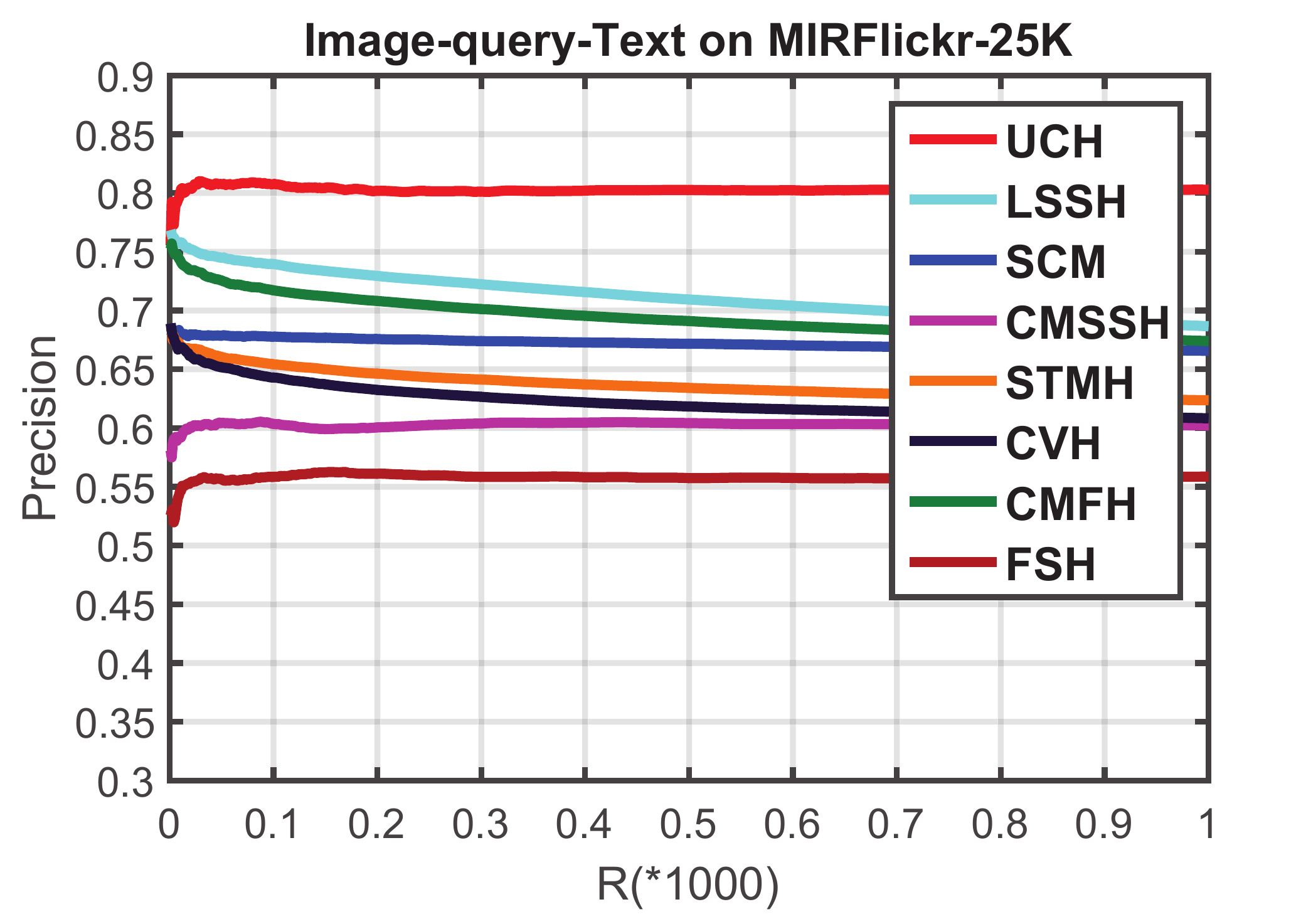}
			
	\end{minipage}}
	\subfloat[]{
		\begin{minipage}[b]{.33\linewidth}
			\centering
			\includegraphics[width=.99\textwidth]{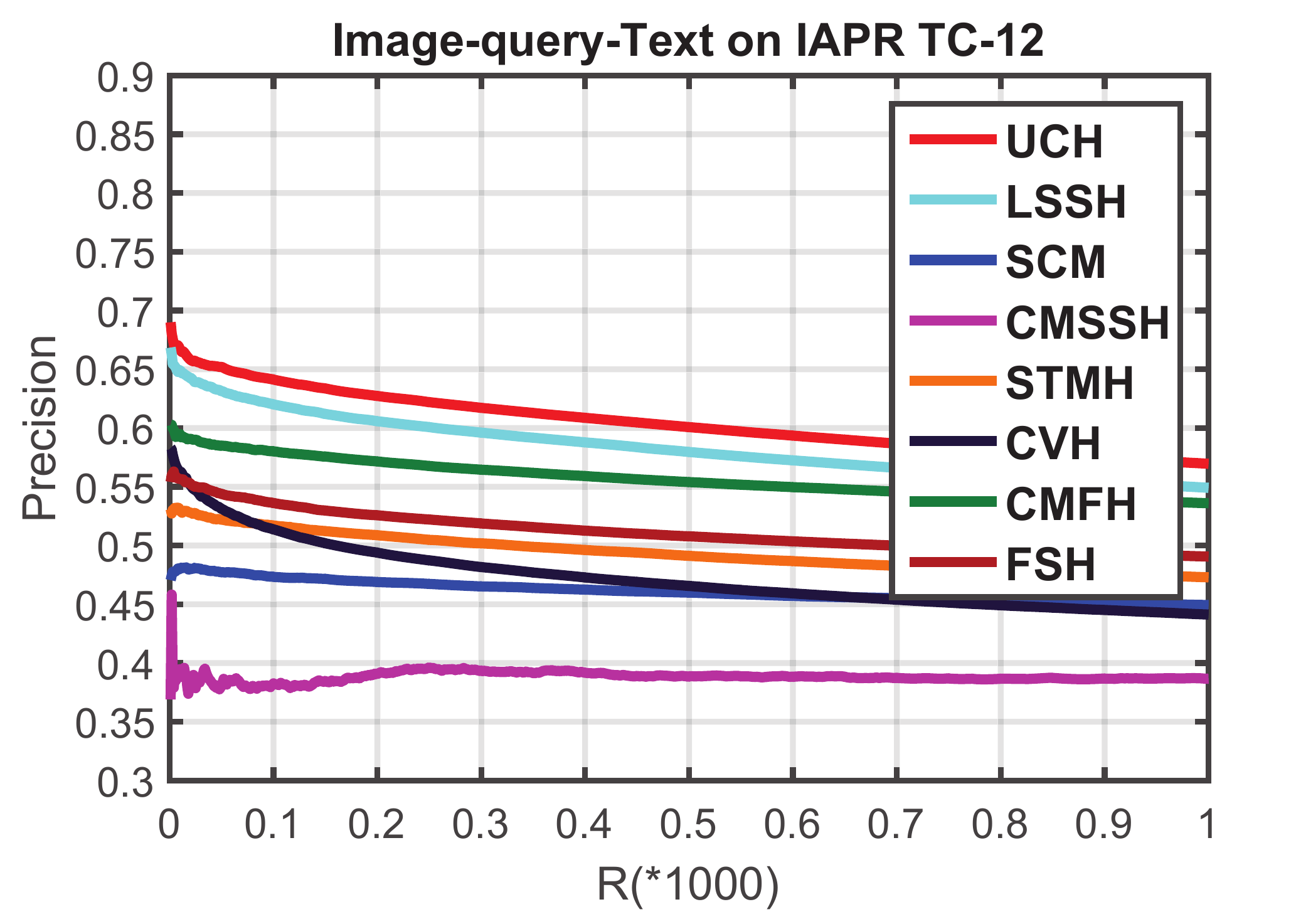}
			
	\end{minipage}}
	\subfloat[]{
		\begin{minipage}[b]{.33\linewidth}
			\centering
			\includegraphics[width=.99\textwidth]{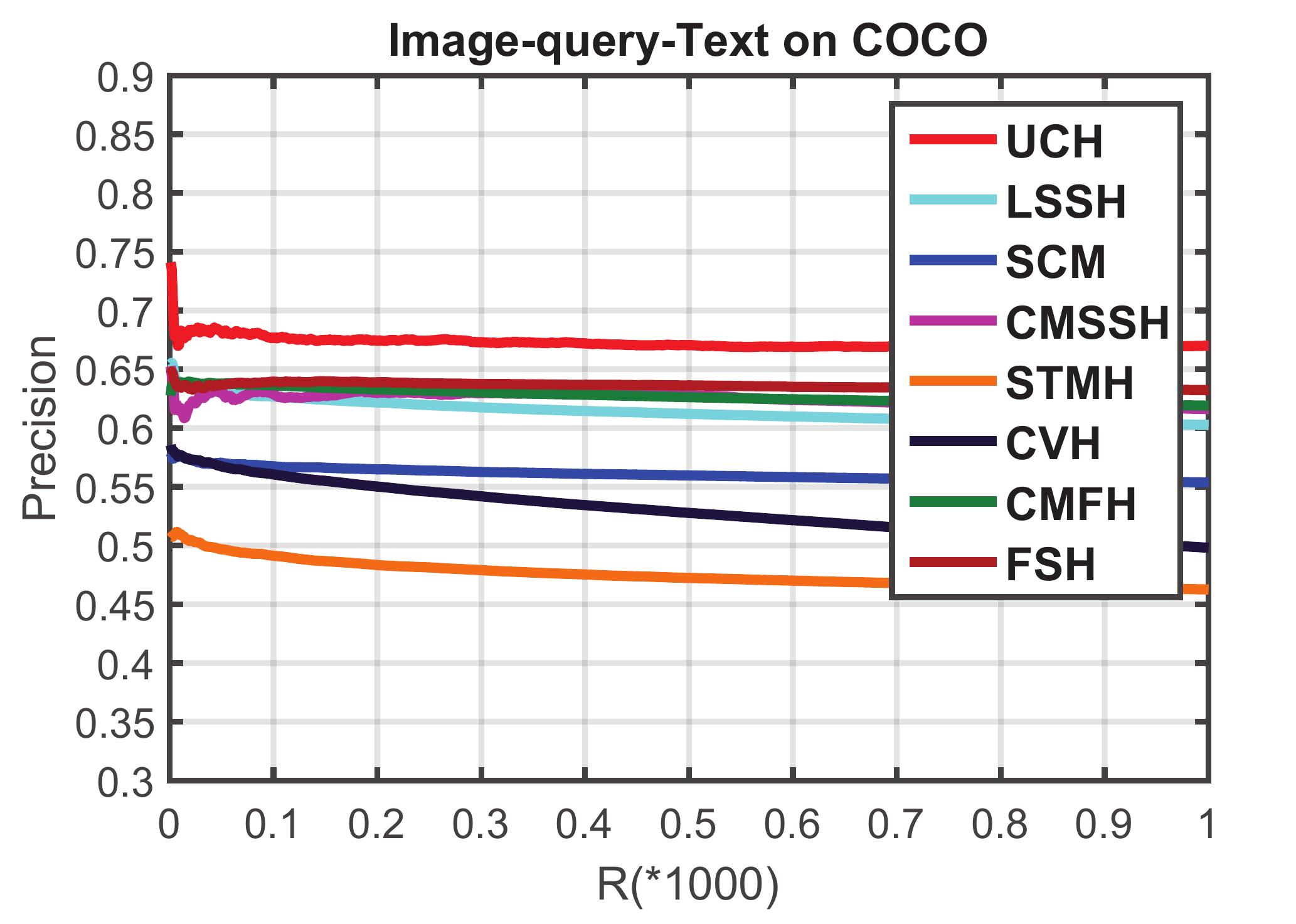}
			
	\end{minipage}}\\
	\subfloat[]{
		\begin{minipage}[b]{.33\linewidth}
			\centering
			\includegraphics[width=.99\textwidth]{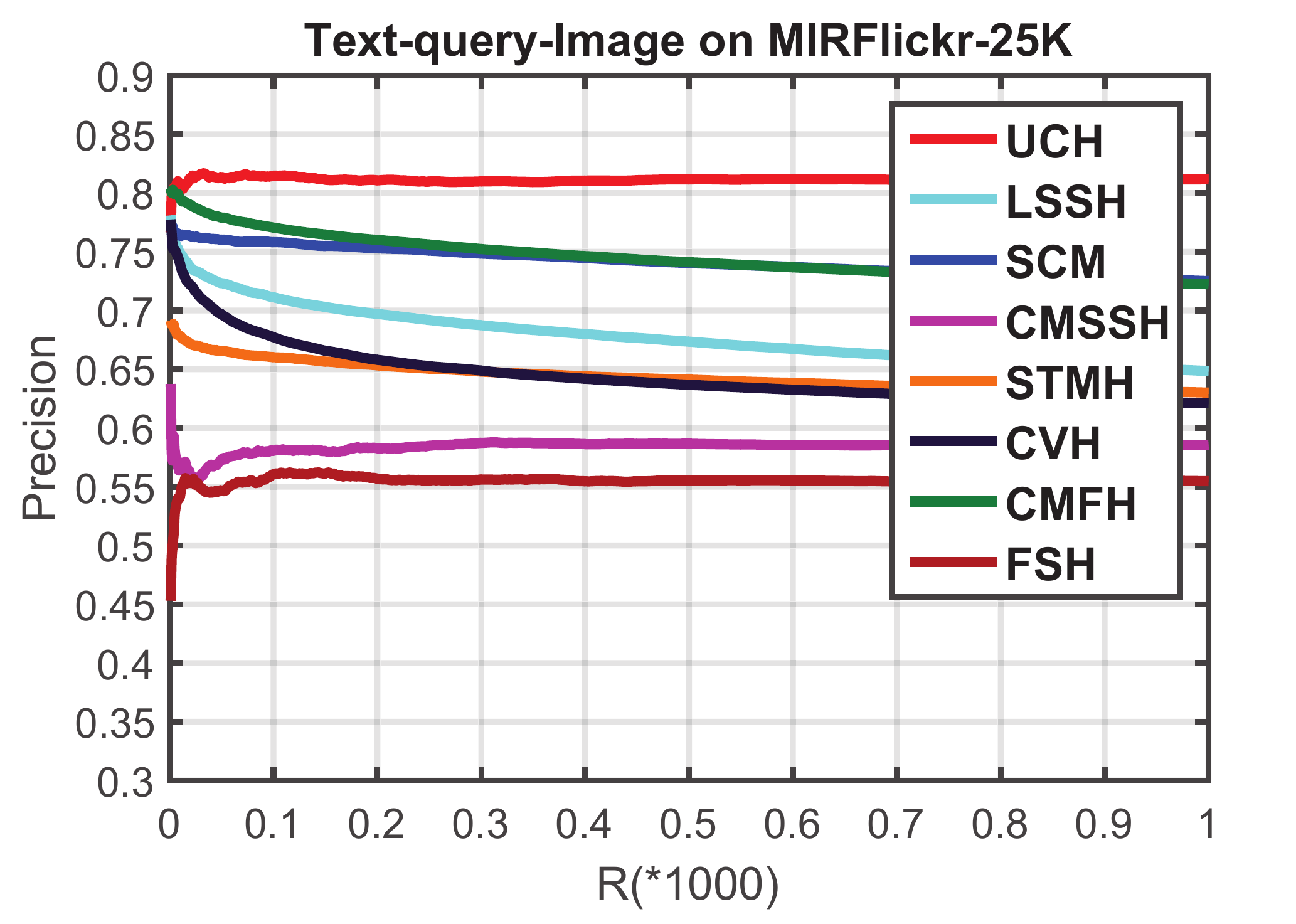}
	\end{minipage}}
	\subfloat[]{
		\begin{minipage}[b]{.33\linewidth}
			\centering
			\includegraphics[width=.99\textwidth]{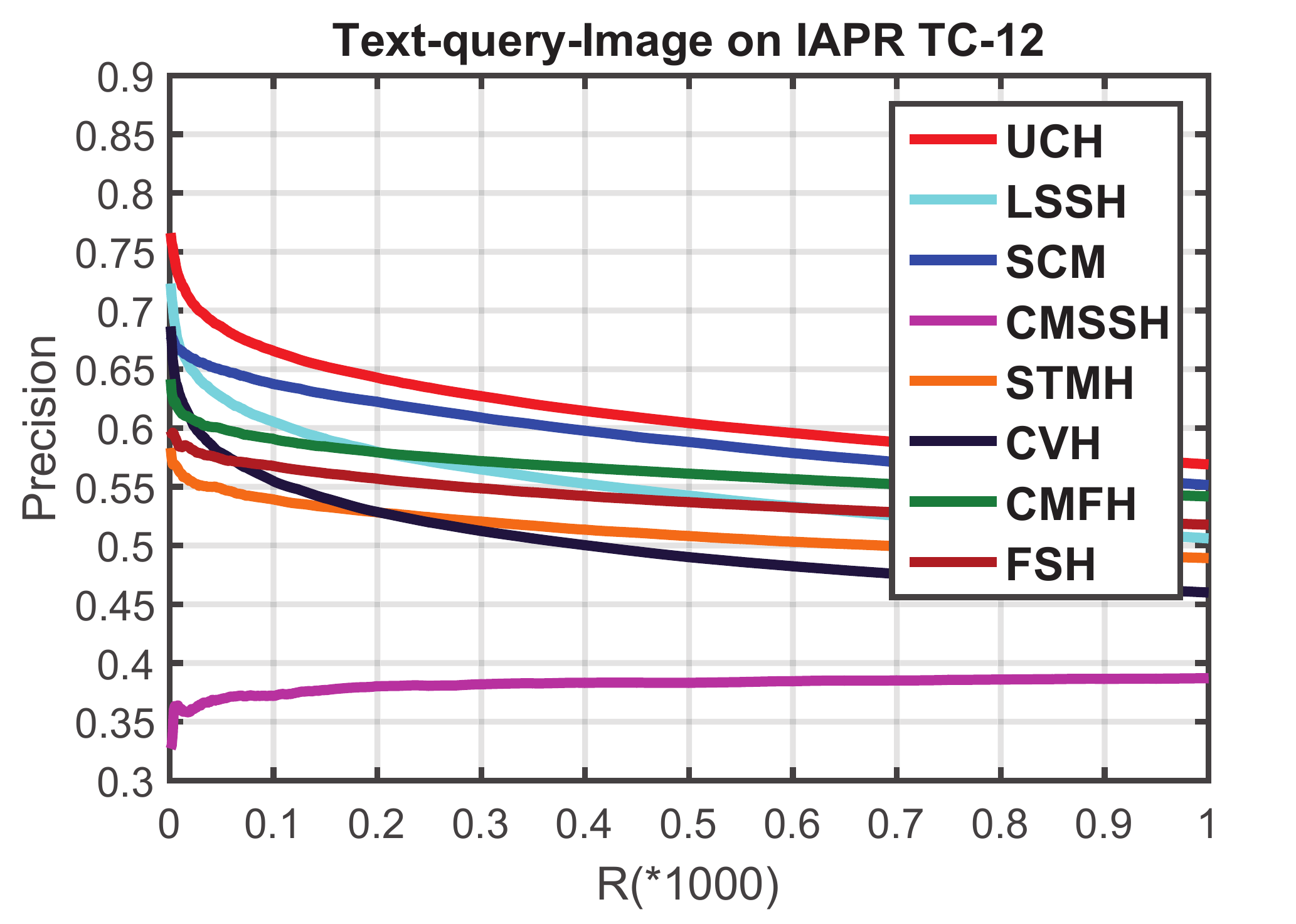}
	\end{minipage}}
	\subfloat[]{
		\begin{minipage}[b]{.33\linewidth}
			\centering
			\includegraphics[width=.99\textwidth]{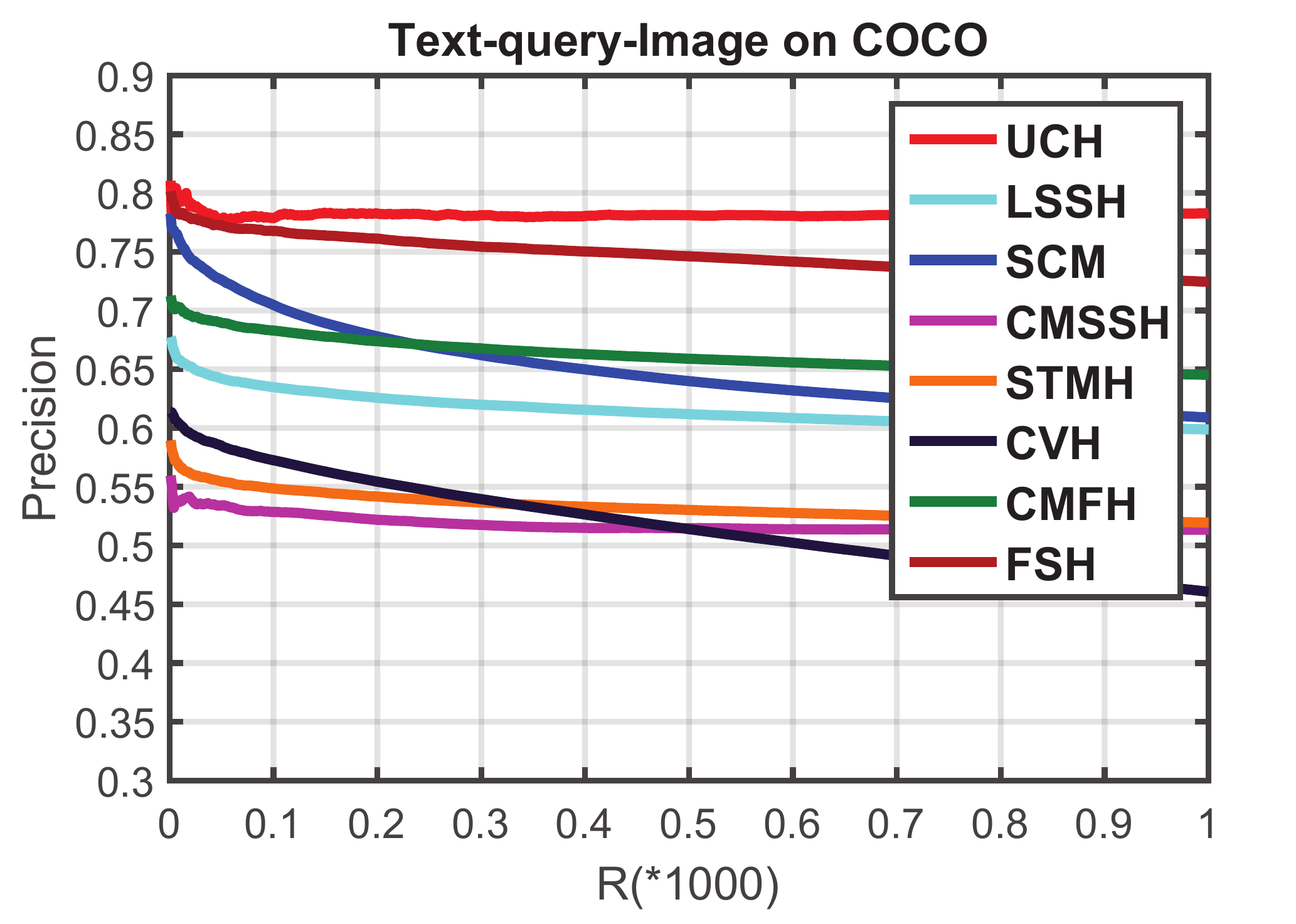}
	\end{minipage}}
	\caption{Precision curves with top 1,000 retrieved instances on MIRFlickr25k, IAPR TC-12, and COCO with 64 bits.}
	\label{fig::Pcurve}
\end{figure*}

\subsection{Experiment Results}
Table~\ref{result::MAP} reports the MAP results for our proposed UCH and the compared the-state-of-art methods on MIRFlickr-25K, IAPR TC-12, and Microsoft COCO datasets. As shown in Table~\ref{result::MAP}, we group these compared methods into two categories: supervised and unsupervised. From Table~\ref{result::MAP}, some conclusions can be obtained: 1) Among the traditional methods, SCM and CMSSH depending on additional supervised information, achieve relatively good performance on both retrieval tasks on average. 2) CMFH, FSH, and LSSH can achieve comparable performance in general. 3) By comparing all these methods, our proposed UCH achieves the highest MAP score at all code lengths and significantly outperforms other competitors. This may be because that almost these compared methods learn common representations and hash codes dividually, which limits their retrieval accuracy. By contrast, the proposed unsupervised coupled cycle deep hashing network unifies common representation learning and hash codes learning together, which can be optimized in one framework. Therefore, more reliable hash codes can be achieved. 

Additionally, we provide the Precision-Recall curves~(PR curves) and Precision of the top 1,000 curves~(Precision\emph{@}1000) in Fig.~\ref{fig::PRcurve} and Fig.~\ref{fig::Pcurve}, respectively. PR curve is obtained by varying the Hamming radius from $0$ to $64$ with a stepsize $1$. Precision\emph{@}1000 presents the precision for the top 1,000 retrieved instances. Fig.~\ref{fig::PRcurve} shows PR curves of all state-of-the-art methods with 64-bit hash codes on three benchmark datasets. From Fig.~\ref{fig::PRcurve} and Fig.~\ref{fig::Pcurve}, similar conclusions mentioned above can be achieved. 
\begin{table}[!t]
	\begin{center}
		\caption{Compared the proposed UCH with UGACH. MAP score evaluated on MIRFlickr25K.}
		\label{result::VsUGACH Flickr}
		\begin{tabular}{|c|c|c|c|c|}
			\hline
			\multirow{2}{*}{Task} &\multirow{2}{*}{Method} &\multicolumn{3}{c|}{MIRFlickr25K}\\
			\cline{3-5}
			& & 16 & 32 & 64 \\
			\hline
			\multirow{2}{*}{Image$\rightarrow$Text}
			& UGACH          & 0.603 & 0.607 & 0.616 \\
			& \textbf{UCH}    & $\mathbf{0.654}$ & $\mathbf{0.669}$ & $\mathbf{0.679}$  \\
			\hline
			\multirow{2}{*}{Text$\rightarrow$Image}
			& UGACH        & 0.590 & 0.632 & 0.642 \\
			\cline{2-5}
			& \textbf{UCH}  & $\mathbf{0.661}$ & $\mathbf{0.667}$ & $\mathbf{0.668}$ \\
			\hline
		\end{tabular}
	\end{center}
\end{table}
\begin{table}[!t]
	\begin{center}
		\caption{Compared the proposed UCH with UGACH. MAP score evaluated on IAPR TC-12.}
		\label{result::VsUGACH IAPR}
		\begin{tabular}{|c|c|c|c|c|}
			\hline
			\multirow{2}{*}{Task} &\multirow{2}{*}{Method} &\multicolumn{3}{c|}{IAPR TC-12}\\
			\cline{3-5}
			& & 16 & 32 & 64 \\
			\hline
			\multirow{2}{*}{Image$\rightarrow$Text}
			& UGACH          & 0.439 & 0.454 & 0.479 \\
			& \textbf{UCH}    &  $\mathbf{0.447}$ & $\mathbf{0.471}$ & $\mathbf{0.485}$ \\
			\hline
			\multirow{2}{*}{Text$\rightarrow$Image}
			& UGACH        & 0.433 & 0.456 & 0.480 \\
			\cline{2-5}
			& \textbf{UCH}  & $\mathbf{0.446}$ & $\mathbf{0.469}$ & $\mathbf{0.488}$\\
			\hline
		\end{tabular}
	\end{center}

\end{table}

\textbf{Comparison UCH with UGACH.} We additionally compare our proposed UCH with UGACH, which is a representative deep learning based method proposed recently. For fair comparison, the same CNN-F network is adopted to extract deep features for UGACH. Table~\ref{result::VsUGACH Flickr} and Table~\ref{result::VsUGACH IAPR} show the results of comparison between UCH and UGACH in term of MAP values on MIRFlickr and IAPR TC-12 datasets. It is obvious that our proposed UCH outperforms UGACH with different code lengths in all cases. The main reason may be that UGACH just calculates relationship once among instances with deep features in the data preprocessing rather than update it iteratively, which causes that the built correlation is lack of accuracy and thus the retrieval performance is constrained. With no need to build similarity matrix, our UCH exploiting modality correlation by generating modality data bi-directionally between two different modalities can learn more powerful representations. Therefore, more reliable hash codes can be achieved with the proposed UCH.

\section{Conclusions}
In this paper, we proposed a novel unsupervised coupled cycle generative adversarial hashing network, for large-scale cross-modal retrieval. The uniqueness of our method is that powerful common representations and reliable hash codes can be learned in an unified framework without using any label information. Moreover, common representation learning and hashing learning, interacting with a coupled manner, can achieve optimal performance at the same time. The extensive experiments on three widely-used datasets show that our proposed model achieves state-of-the-art performance in cross-modal retrieval tasks.

\section{ Acknowledgments}
Our work was supported in part by the National Natural Science Foundation of China under Grant 61572388 and 61703327, in part by the Key R\&D Program-The Key Industry Innovation Chain of Shaanxi under Grant 2017ZDCXL-GY-05-04-02, 2017ZDCXL-GY-05-02 and 2018ZDXM-GY-176, and in part by the National Key R\&D Program of China under Grant 2017YFE0104100.

\bigskip
\bibliographystyle{aaai} 
\bibliography{AAAI-LiC.4747}
\end{document}